\documentstyle[twocolumn,aps]{revtex}
\begin{document}
\draft
\title{\bf {Magnetization of mesoscopic superconducting discs}}
\author{P. Singha Deo\cite{eml}, V. A. Schweigert \cite{sc}, and F.
M.  Peeters \cite{fmp}}
\address{Department of Physics, University of Antwerp (UIA),
B-2610 Antwerpen, Belgium.}
\author{and\\A. K. Geim}
\address{High Field Magnet Laboratory, University of Nijmegen, 6525
ED Nijmegen, the Netherlands}
\maketitle
\begin{abstract}
Solutions of Ginzburg-Landau eqns. coupled with three dimensional
Maxwell eqns. reveal intriguing magnetic response of small
superconducting particles, qualitatively different from the two
dimensional approximation but in agreement with recent experiments.
Depending on the radius and thickness first or second order
transitions are found for the normal to superconducting state.  For a
sufficient large radius of the disc several transitions in the
superconducting phase are obtained which correspond to different
angular momentum giant vortex states.  The incorporation of the
finite thickness in the calculation is crucial in order to obtain
agreement with the position and the size of these jumps, and the line
shape and magnitude of the magnetization curves.
\end{abstract}
\narrowtext
%\newpage

Recent advances in micro-fabrication technology and measurement
techniques have allowed the first studies of thermodynamic properties
of well controlled mesoscopic superconducting particles
\cite{bui,mos,gei}. The samples are mesoscopic in the sense that
their size is comparable to the Ginzburg-Landau (GL) coherence
length.  Buisson {\it et al} \cite{bui} performed magnetization
measurements on an ensemble of discs with large separation between
them in order to make the dipolar interaction between the discs
negligible.  They found oscillatory behavior in the magnetization
near the superconducting transition temperature and showed that the
linearized Ginzburg-Landau (LGL) eqn. are able to explain
qualitatively part of their experiments but there are some major
discrepancies in size and position of the jumps in the magnetization.
Recently Geim {\it et al} \cite{gei} used sub-micron Hall probes to
detect the magnetization of {\it single} superconducting discs with
size down to 0.1 $\mu m$.  At different applied fields the discs show
various kinds of phase transitions within the superconducting state
and between the superconducting and normal state which can be first
or second order depending on the sample dimensions and temperature.
The aim of this letter is to explain this intriguing behavior and to
give a quantitative analysis of these magnetization experiments.

A number of earlier works studied: 1) disc geometries in the
framework of the LGL eqn. where a uniform magnetic field is assumed
\cite{bui,mos,zw,5} in the discs, or 2) cylindrical geometries in the
framework of the non-linear GL eqns. \cite{mos1}. The first type of
approximations is reasonable if one is interested in the
superconducting-normal state boundary where the order parameter is
very small and the magnetic field equals the external one. The second
type of approach is also valid away from this boundary but does not
work for discs of finite thickness. Here we are interested to explain
the full magnetization curve as a function of the external magnetic
field for mesoscopic discs with finite thickness. Magnetization is
the ideal tool to understand the superconducting state deep inside
the phase boundary.

It is known that a type II superconducting cylinder in a magnetic
field parallel to its axis can exist in three phases of
superconductivity.  Below $H_{c1}$ we have a pure superconducting
state, between $H_{c1}$ and $H_{c2}$ there is the mixed state and
between $H_{c2}$ and $H_{c3}$ we have surface superconductivity or
the giant vortex state \cite{deg}.  As the height of the cylinder is
reduced, so that it becomes a disc, the magnetic field penetration
into the sample is determined by the penetration length $\lambda$ as
well as the disc thickness, due to geometrical form factors. This
makes the above simple divisions no longer applicable. Our problem
requires a 3D solution, instead of their 2D version which turns out
to be essential in order to understand the experiments of Refs.
\cite{bui} and \cite{gei}.

We consider a superconducting $Al$ discs ($\kappa$=.28) with radius
$R$ and thickness $d$ immersed in an insulating medium. Thin film
discs are known to behave as type II samples \cite{10} and can be
described by the GL theory \cite{bui}. For mesoscopic $Al$ samples
(squares and thin wires), the GL theory has been successfully
employed to explain the phase boundary \cite{mos}.  Hence as a first
approximation, neglecting the non-local effects, we solve the system
of two coupled GL eqns.
\begin{equation}
\label{eq1}
\frac{1}{2m}\left(-i\hbar\vec \nabla -\frac{2e\vec A}{c}\right)^2\Psi=
-\alpha\Psi-\beta\Psi|\Psi|^2,
\end{equation}
\begin{equation}
\label{eq2}
\vec \nabla \times \vec \nabla\times \vec A=\frac{4\pi}{c}\vec j,
\end{equation}
where the density of superconducting current $\vec j$ is given by
\begin{equation}
\label{eq3}
\vec j=\frac{e\hbar}{im}
\left(\Psi^*\vec \nabla \Psi-\Psi \vec\nabla\Psi^*\right)
-\frac{4e^2}{mc}|\Psi|^2\vec A.
\end{equation}
The boundary conditions on the disc corresponding to zero current
density in the insulator medium is
\begin{equation}
\label{eq4}
\left(-i\hbar\vec\nabla-\frac{2e\vec A}{c}\right)\Psi \mid_n=0,
\end{equation}
where the subscript $n$ denotes the component normal to the disc
surface.  The boundary condition for the vector potential is such
that far away from the superconducting disc the field equals the
applied field $\vec H=(0,0,H_0)$, i.e., $\vec A|_{\vec
\rho\rightarrow \infty}=\vec e_{\phi}H_0\rho /2$. Here $\vec
e_{\phi}$ denotes the azimuthal direction and $\rho$ is the radial
distance from the disc center.

Using dimensionless variables and the London gauge $div \vec A=0$, we
rewrite the system of eqns. (\ref{eq1}-\ref{eq4}) into the following
form
\begin{equation}
\label{eqn1}
\left(-i\vec \nabla -\vec A\right)^2\Psi=
\Psi (1-|\Psi|^2),
\end{equation}
\begin{equation}
\label{eqn2}
-\kappa^2\triangle \vec A=
\frac{1}{2i}
\left(\Psi^*\vec \nabla \Psi-\Psi \vec\nabla\Psi^*\right)-|\Psi |^2\vec A.
\end{equation}
Here the distance is measured in units of the coherence length
$\xi=\hbar/\sqrt{-2m\alpha}$, the order parameter in $\psi
_0=\sqrt{-\alpha/\beta}$, the vector potential in $c\hbar/2e\xi$,
$\kappa =\lambda /\xi$ is the GL parameter, and
$\lambda=c\sqrt{m/\pi}/4e\psi _0$ is the penetration length. We
measure the magnetic field in $H_{c2}=c\hbar/2e\xi ^2=\kappa
\sqrt{2}H_c$, where $H_c=\sqrt{-4\pi \alpha/\beta}$ is the critical
field. The difference of the Gibbs free energy $G$ between the
superconducting and the normal state measured  in $H_c^2V/8\pi$ can
be expressed through the integral
\begin{equation}
\label{free}
G=\int \left( 2(\vec A-\vec A_0)\vec j-|\Psi|^4\right)d\vec r/V,
\end{equation} 
over the disc volume $V=\pi R^2d$, where $\vec A_0=\vec
e_{\phi}H_0\rho/2$ is the external vector potential, and $\vec
j=(\Psi^*\vec \nabla \Psi-\Psi \vec\nabla\Psi^*)/2i -|\Psi |^2\vec A$
is the dimensionless superconducting current.

For thin discs, the magnetic field is uniformly distributed along the
$z-$direction. When the disc thickness becomes comparable to the
penetration length, the magnetic field is expelled from the disc due
to the Meissner effect. The field penetrates only a depth $\lambda$
inside the disc. Therefore, the variation of the vector potential in
the direction parallel to the applied field becomes rather strong for
$d>\lambda$.  Nevertheless, this does not lead to essential
variations of the order parameter in this direction, in discs thinner
than the coherence length.  Representing the order parameter as a
series $\Psi(z,\vec\rho)=\sum_k cos(k\pi z/d)\Psi _k(\vec \rho),$
which obeys the boundary condition (\ref{eq4}) at the disc sides
$z=\pm d/2$ and using the first GL eqn. (\ref{eq1}), one can
verify that the part of the order parameter which is uniform in the
z-direction i.e., $\Psi_0$ gives the main contribution to the
expansion for $(\pi \xi/d)^2\gg 1$. Therefore, we may assume a
uniform order parameter along the $z-$direction and average the first
GL eqn. over the disc thickness. Since the order parameter does
not change in the $z-$direction, both the superconducting current and
the vector potential have no $z-$component. Then the boundary
condition (\ref{eq4}) is automatically fulfilled on the upper and
lower disc sides.

Our 3D calculations show that the  discs studied experimentally in
Ref. [1] exist in a regime of surface superconductivity, or the giant
vortex state i.e., between the $H_{c2}$ line and the $H_{c3}$ line.
If the thickness of the discs is further reduced then we go over to
the mixed regime and the giant vortex breaks up into many vortices if
the radius of the disc is sufficiently large, even for a type I
sample.

Therefore, we consider the situation with a fixed value of the
angular  momentum $L$ for the order parameter $\Psi(\vec
\rho)=F(r)exp(iL\phi)$, when both the vector potential and the
superconducting current are directed along $\vec e_{\phi}$. Then
eqns.  (\ref{eqn1}-\ref{eqn2}) are reduced to the following form
\begin{equation}
\label{eqnn1}
-\frac{1}{\rho}\frac{\partial }{\partial \rho}\rho
\frac{\partial F}{\partial \rho}
+(\frac{L^2}{\rho^2}-2\frac{L}{\rho}<A>+<A^2>)F=F(1-F^2),
\end{equation}
\begin{equation}
\label{eqnn2}
-\kappa^2\left(\frac{\partial}{\partial \rho}\frac{1}{\rho}
\frac{\partial \rho A}{\partial \rho}
+\frac{\partial ^2A}{\partial z^2}\right)
=\left((\frac{L}{\rho}-A)F^2\right)\theta(r/R)\theta(2|z|/d),
\end{equation}
where $\theta(x<1)=1$, $\theta(x>1)=0$; $\vec A=\vec e_{\phi}A$; $R$,
$d$ are the dimensionless disc radius and thickness, respectively;
the brackets $<>$ means averaging over the disc thickness
$<f(\rho)>=\int_{-d/2}^{d/2}f(z,\rho)dz/d$.

The magnetic field created by the superconducting current in the disc
decreases in strength away from the disc as a magnetic dipole field:
$H\sim 1/r^3$.  Because of this, in our numerical calculations the
condition for the vector potential is transferred from infinity to
the boundaries of the simulation region as follows:
$A(z,r=R_s)=\frac{1}{2}H_0R_s, \quad
A(|z|=d_s,\rho)=\frac{1}{2}H_0\rho,$ where $R_s,d_s\gg R,d$ are the
sizes of the simulation region in the radial and $z$ direction,
respectively. We use typically $R_s,d_s=(5\div 10)max(R,d)$, and
checked that an increase of the simulation region does not change our
results substantially.  The boundary conditions for the order
parameter
\begin{equation}
\label{bound2}
\frac {\partial F}{\partial \rho}|_{\rho=R}=0,\quad
\rho\frac {\partial F}{\partial \rho}|_{r=0}=0, \quad
\end{equation}
correspond to zero current density at the disc surface and  a finite
value of the first derivative of $F$ at the disc center.  To solve
numerically the system of eqns. (\ref{eqnn1}-\ref{eqnn2}) we apply a
finite-difference representation of the GL and 3D Maxwell eqns.  on
the space grid $\rho_i$, $z_j$.

Discs in three different regimes will be considered 1) type II, 2)
type I and 3) multiple type I behavior.  When we compare the
theoretical results with the experimental data we have to keep in
mind that experimentally the magnetization will depend on the filling
fraction of the Hall bars used as detectors which is not exactly
known. Also because of the square geometry of the Hall detector whose
sides are of the same size as the diameter of the largest disc it
will underestimate the magnetization (the flux expelled) of the
smaller discs. These effects will result in an unknown scale factor
for the magnetization of order 1. In order to have a comparison of
relative magnitudes such as the size of the jumps in magnetization,
we scale the theoretical results such that they have the same maximum
magnetization as observed experimentally.  When determining the
magnetization from the LGL eqn., the same method as in Ref.
\cite{bui} was used; we have assumed the Abrikosov parameter $\beta$
to be 1.0 as done in Ref. \cite{bui}.  We compare our theoretical
results with the experimental results on $Al$ discs at 0.4 K of Geim
{\it et al} \cite{gei} and took for the zero temperature coherence
length $\xi(0)$= 250 $nm$ and the penetration length $\lambda(0)$=70
$nm$ as estimated in Ref. \cite{gei}. The disc thickness and radius
are also given in Ref. \cite{gei} and therefore our theory does not
contain any fitting parameters.

Fig$.\,1$ shows the magnetization curves for an Al disc of thickness
d=0.15 $\mu m$ and radius R=0.315 $\mu$m. Large solid dots are
the experimental data and exhibit a continuous
superconducting-normal transition; the dotted curve is the solution
from the LGL eqn. whereas the thin solid curve is the numerical
solution of the non-linear GL eqns. coupled to the three
dimensional Maxwell eqn. The dotted curve is scaled by 0.158
and the solid curve by 0.537. Surprisingly the dotted curve gives a
line shape in closer agreement with the experiment but its magnitude
is clearly too large.  There is some improvement in the line shape
(dashed curve) if we reduce the disc thickness to 0.07 $\mu m$ in
which case the radius of the disc was changed to 0.31 $\mu m$ in
order to keep the critical field the same (the magnetization was
scaled by 1).  Therefore, we suspect that the effective thickness 
of the
disc which is still superconducting is much smaller than the actual
thickness.

Fig$.\,2$ shows the magnetization curves for a larger disc of
thickness 0.15 $\mu$m and radius 0.473 $\mu$m at 0.4K. The same
symbol and curve conventions are used as in previous figure. The
dotted curve is scaled by 0.124 and the thin solid curve by 0.581. It
is obvious that the dotted curve is very different in shape and
magnitude and shows a jump in magnetization at a very different value
compared to the experimental curve. This clearly demonstrates that a
LGL eqn. with a homogeneous magnetic field distribution over the disc
is not appropriate in this case. The finite thickness of the disc
results in very important geometrical corrections to the field
profile which influences the superconducting state appreciably. Note
that the magnetic field dependence of the experimental curve is well
described by the thin solid curve: 1) the slope of the magnetization
curve, 2) the non-linear behavior near the step in the magnetization,
and 3) also the magnetic field at which the step in the magnetization
takes place is correctly predicted. The experiment shows a first
order transition from the superconducting to normal state at a
magnetic field of 70.86 G while our calculations still predict a
transition to the $L=1$ superconducting state which becomes normal at
81.5 G. The origin of this small discrepancy is still not clear to us
but may be due to effects of disorder \cite{6}.

The magnetization curves for a disc of thickness 0.15 $\mu$m and
radius 1.2 $\mu$m, is shown in Fig$.\,3$.  The symbol and the curve
conventions are again the same as before. The dotted curve is scaled
by 0.062 and the thin solid curve by 0.775. Note that the LGL eqn. in
this case gives the same type of discrepancies as found in the
experiment of Buisson {\it et al} \cite{bui}.  Firstly they found
that the magnitude of the jumps in the magnetization as obtained from
a solution of the LGL eqn., are too large compared to the
experimental results.  Buisson {\it et al} argued that this was due
to an ensemble averaging in their experiment.  The single disc
experiment of Geim {\it et al} rules out this possibility. It is true
that the magnitude of the jumps in the single disc experiment is much
larger than in the many disc experiment, but still, for the single
disc, the jumps are much smaller than those obtained from a solution
of the LGL eqn (compare the dotted curve with the experimental data
in Fig$.\,3$).  Our thin solid curve gives precisely the same
magnitude for the jumps in the magnetization as in the experiment and
also the correct magnitude of magnetization and magnetic field for
most of the transitions.  Secondly, they found that the position of
the first jump in the magnetization obtained from the LGL eqn. is
much below that of the experimental curve. No proper explanation
could be given for this.  Our dotted curve gives similar discrepancy
of approximately the same magnitude and the thin solid curve settles
this dispute.  The first jump coincides with that of the experimental
curve.  We find that if we keep the radius unchanged and decrease the
disc thickness then the upper critical field and the number of jumps
in the magnetization curve remain unchanged. Only the position of the
first peak shifts towards lower magnetic field which will of course
also lead to an increase in the magnetic field spacing at which the
jumps in magnetization occurs.  This obviously is due to the fact
that as the disc thickness is reduced the magnetic field inside the
disc increases at a faster rate and so the transition to the first
fluxoid state occurs at lower applied fields. Near the critical field
of course the field inside does not depend on the thickness and is
the same as the external field. This once again shows that the
de-magnetization factor is crucial for these discs and they determine
largely the shape of the magnetization curves.  Another interesting
point  to be noted is that the experimental curve shows a gradually
decreasing interval of magnetic field at which the jumps occur. It
was shown in Ref. \cite{zw} that within the LGL theory flux
quantization condition does not imply that the jumps in the
magnetization will occur at regular intervals. As long as the order
parameter at the central region of the disc is not negligible, the
interval will decrease slowly and hence only for large values of $L$
the interval will be the same as that given by the flux quantization
condition. But our 3D solution show a more drastic decrease in the
interval (since the first few jumps do not coincide with the LGL
theory whereas the last jump does). The reason is that at smaller
fields the field inside the disc changes very slowly.  In fact there
is even a small regime where the field at the center of the disc can
decrease with increasing applied field (see curves for $L$=1 and
$L$=2 in the inset of Fig$.\,3$). But at high fields the field inside
the disc increases almost in the same way as the external field. In
the inset of Fig$.\,3$ we have plotted the magnetic field
distribution $H(\rho,z$=0,$\theta$=0) of the system considered in
Fig$.\,3$ for ten values of the applied field. The corresponding
value of the angular momentum of the equilibrium superconducting
state and the magnetic field is also indicated. For $L\ne0$ there is
substantial penetration of the magnetic field in the center of the
disc while a ring like region near the edge of the disc remains
superconducting.  This is the well known surface superconductivity or
the giant vortex state that arises in a finite sample due to the
lower energy of the edge states as compared to the bulk states
\cite{deg}. As soon as the system becomes ring like two equal and
opposite screening currents flow in the system (one along the outer
radius of the ring like region and the other along the inner one).
Note that for the highest field shown in the inset of Fig$.\,3$ there
is no bulk superconductivity and only a tiny region of surface
conductivity survives which is responsible for the dip in the
magnetic field near the edge of the disc.

All along we have assumed that the system evolves along the free
energy minimum and obtained quantitative agreement for the position,
magnitude and periodicity of the jumps in the magnetization as well
as the absolute value of the magnitude of magnetization. But there is
one discrepancy.  Note that the theoretical curves in Fig$.\, 3$ show
a critical field which is appreciably smaller than found
experimentally and the total number of jumps in the experiment is 19
compared to 11 in the theory. When we enlarge the disc radius from
1.2 $\mu m$ to 1.57 $\mu m$, the number of jumps in the magnetization
curve is increased to 19 but the upper critical field is reduced to
63 G.  It is also to be noted that the slope of the experimental
curve decreases with increasing field and tends to become parallel to
the field axis at higher magnetic fields.  In finite systems the
Bean-Livingston barrier \cite{bea} at the surface can cause the
system not to evolve along the free energy minimum.  It leads to
jumps in the magnetization at a much larger value of the magnetic
field compared to the value at which the L=1 state becomes the ground
state. We have to decrease the width of the disc to an unreasonable
value of 0.06 $\mu m$ in order to have the first jump at the same
position as that seen in the experiment for the same disc radius.
But, as discussed before, such a decrease in thickness does not
increase the critical field and cannot explain the high field
discrepancy. It is also known that surface defects can destroy the
Bean-Livingston barrier for increasing fields. The previously
mentioned disorder effects \cite{6} may be responsible  for the
increase of the upper critical field but at present no explicit
calculation exists with which we can compare and therefore this is
still an open issue.

This work is supported by the Flemish Science Foundation(FWO-Vl)
grant No: G.0232.96, the European INTAS-93-1495-ext project, and the
Belgian Inter-University Attraction Poles (IUAP-VI).  One of us (PSD)
is supported by a scholarship from the University of Antwerp and FMP
is a Research Director with FWO-Vl.

%\vfill
%\eject
\centerline{FIGURE CAPTIONS}
\noindent FIG$.\,1$.  Magnetization versus the external magnetic
field for a superconducting disc of radius R= 0.315 $\mu m$ and
thickness d=0.15 $\mu m$ at T=0.4 K. Solid dots are the experimental
data from Ref. \cite{gei}. We show the results of the GL theory
including the 3D Maxwell eqn. (solid curve, and dashed curve for
R=0.31 $\mu m$ and d=0.07 $\mu m$), and the result for the LGL-theory
(dotted curve).

\noindent FIG$.\,2$.  The same as Fig$.\,1$ but now for a disc with
radius R=0.473 $\mu m$ and thickness d=0.15 $\mu m$.

\noindent FIG$.\,3$. The same as Fig$.\,1$ for a disc of radius R=1.2
$\mu m$ and thickness d=$0.15$ $\mu m$. In the inset we show the
field distribution in the plane through the center of
the disc for different values of the external magnetic field. Far
away from the center of the disc the magnetic field equals the
external magnetic field.

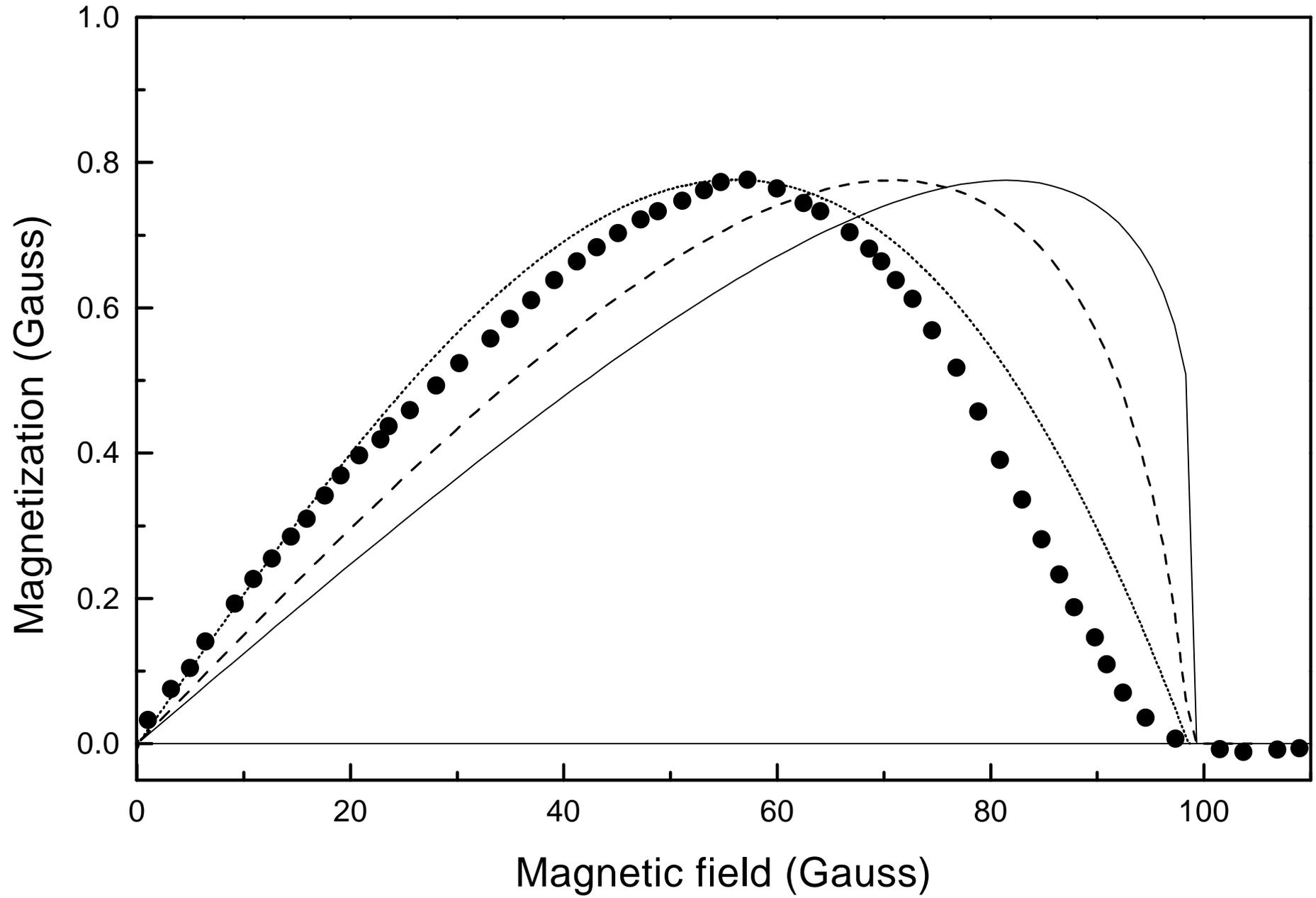
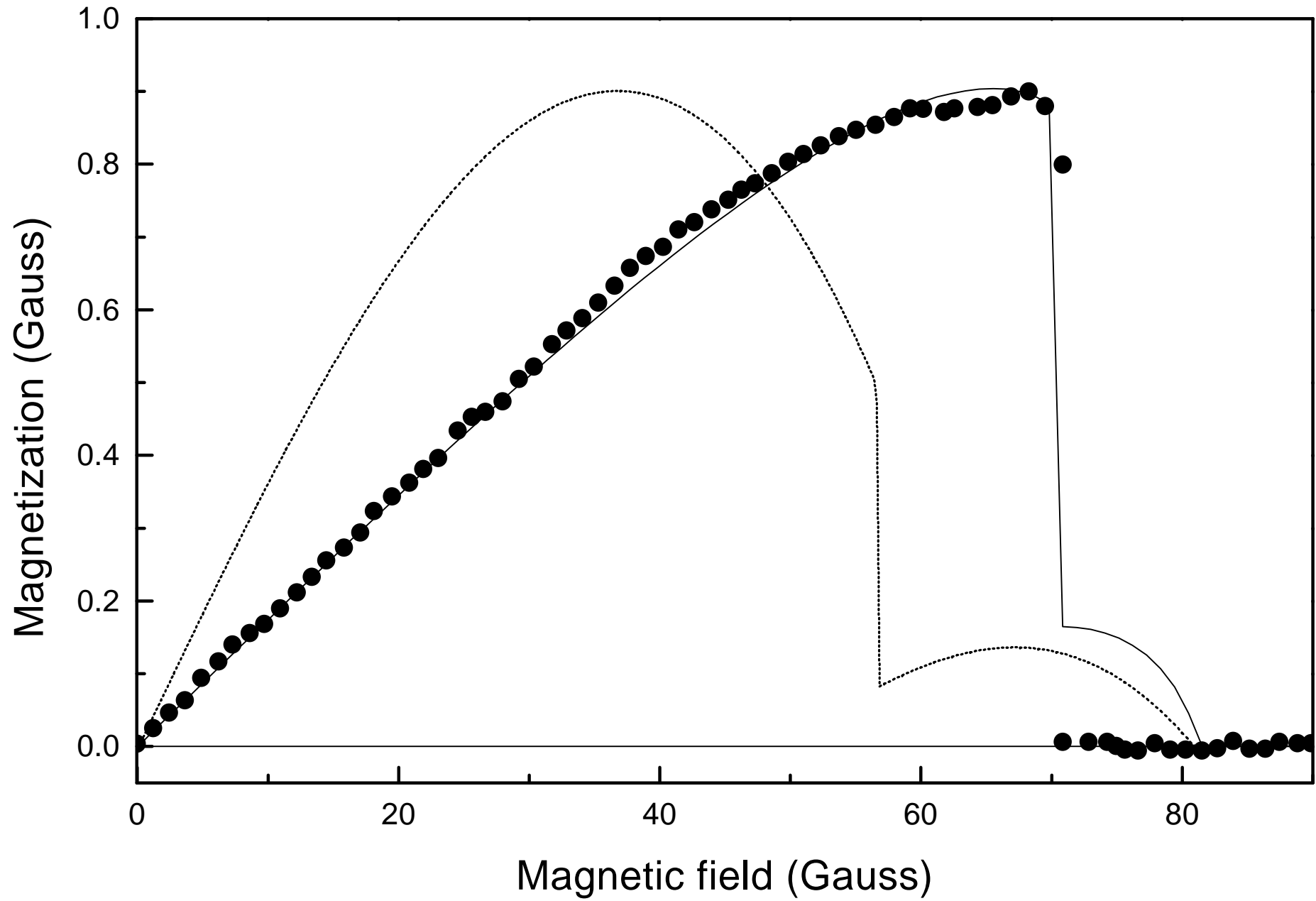
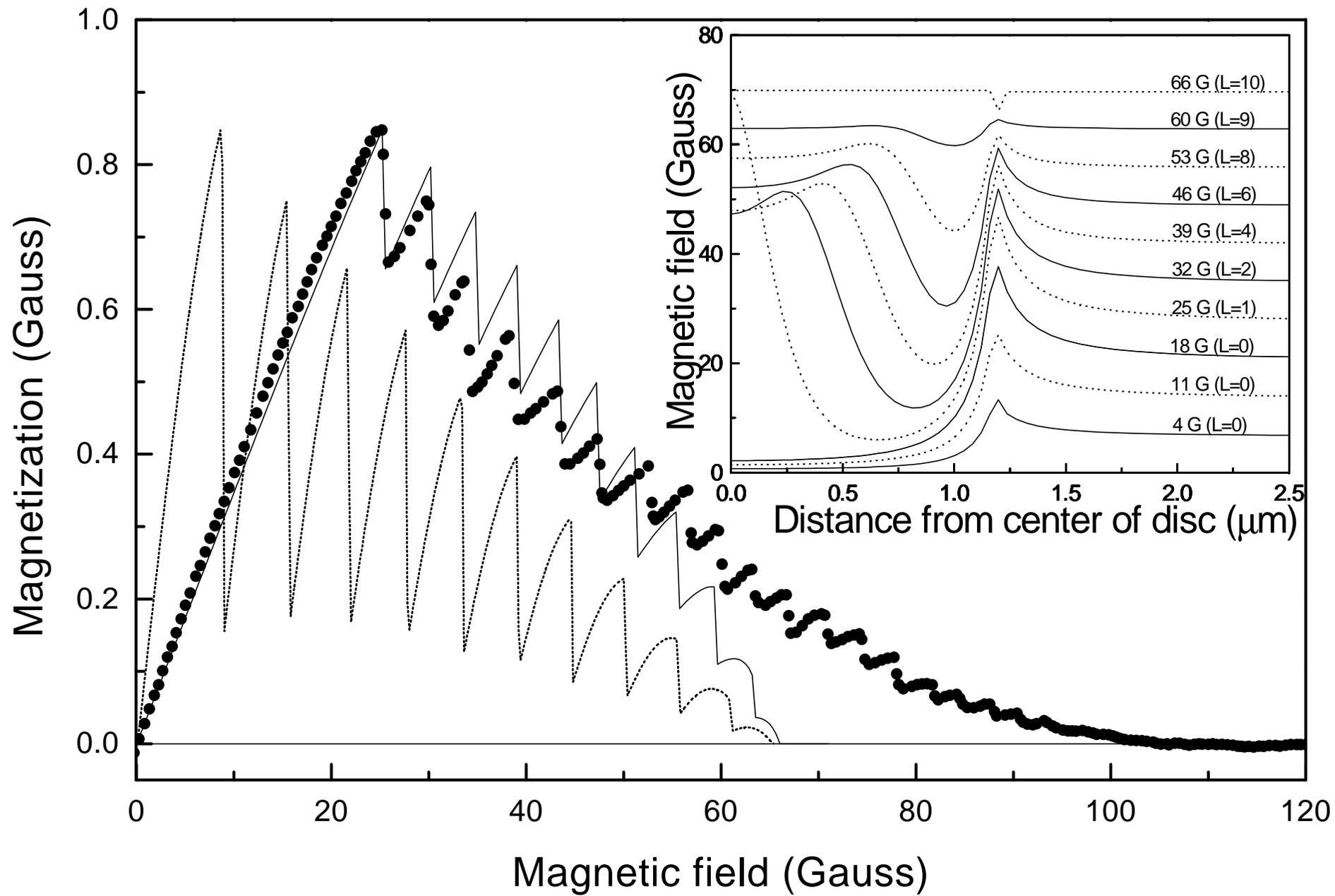
\end{document}